\documentclass[%
 aip,
 amsmath,amssymb,
 reprint,%
]{revtex4-1}

\usepackage{graphicx}% Include figure files
\usepackage{dcolumn}% Align table columns on decimal point
\usepackage{bm}% bold math
\usepackage[utf8]{inputenc}
\usepackage[T1]{fontenc}
\usepackage{mathptmx}
\usepackage{etoolbox}

\usepackage{xcolor}
\usepackage[colorlinks=true, citecolor=blue, linkcolor=blue, urlcolor=blue]{hyperref}
\usepackage{booktabs}
\usepackage{multirow}
\usepackage{tabularx}
\usepackage{array}

\makeatletter
\def\@email#1#2{%
 \endgroup
 \patchcmd{\titleblock@produce}
  {\frontmatter@RRAPformat}
  {\frontmatter@RRAPformat{\produce@RRAP{*#1\href{mailto:#2}{#2}}}\frontmatter@RRAPformat}
  {}{}
}%
\makeatother
\begin{document}

\preprint{AIP/123-QED}

\title[Simulation of complex DNA damage enhancement and biological effect validation for Proton-CAT]{Simulation of complex DNA damage enhancement and biological effect validation for Proton-CAT}
% Force line breaks with \\
\author{Lang Dong}
\affiliation{Key Laboratory of Radiation Physics and Technology of Ministry of Education, Institute of Nuclear Science and Technology, Sichuan University, Chengdu 610064, China}

\author{Dechao An}
\affiliation{Key Laboratory of Radiation Physics and Technology of Ministry of Education, Institute of Nuclear Science and Technology, Sichuan University, Chengdu 610064, China}

\author{Junxiang Wu}
\affiliation{Key Laboratory of Radiation Physics and Technology of Ministry of Education, Institute of Nuclear Science and Technology, Sichuan University, Chengdu 610064, China}
\affiliation{Sichuan Cancer Hospital \& Institute, Sichuan Cancer Center, School of Medicine, University of Electronic Science and Technology of China, Radiation Oncology Key Laboratory of Sichuan Province, Chengdu 610041, China}

\author{Tianle Wang}
\affiliation{Key Laboratory of Radiation Physics and Technology of Ministry of Education, Institute of Nuclear Science and Technology, Sichuan University, Chengdu 610064, China}

\author{Zhao Sun}
\affiliation{Key Laboratory of Radiation Physics and Technology of Ministry of Education, Institute of Nuclear Science and Technology, Sichuan University, Chengdu 610064, China}

\author{Jiajun Kang}
\affiliation{Key Laboratory of Radiation Physics and Technology of Ministry of Education, Institute of Nuclear Science and Technology, Sichuan University, Chengdu 610064, China}

\author{Xianliang Wang}
\affiliation{Sichuan Cancer Hospital \& Institute, Sichuan Cancer Center, School of Medicine, University of Electronic Science and Technology of China, Radiation Oncology Key Laboratory of Sichuan Province, Chengdu 610041, China}

\author{Lintao Li}
\affiliation{Sichuan Cancer Hospital \& Institute, Sichuan Cancer Center, School of Medicine, University of Electronic Science and Technology of China, Radiation Oncology Key Laboratory of Sichuan Province, Chengdu 610041, China}

\author{Shun Lu}
\affiliation{Sichuan Cancer Hospital \& Institute, Sichuan Cancer Center, School of Medicine, University of Electronic Science and Technology of China, Radiation Oncology Key Laboratory of Sichuan Province, Chengdu 610041, China}

\author{Tianli Qiu}
\affiliation{Sichuan Cancer Hospital \& Institute, Sichuan Cancer Center, School of Medicine, University of Electronic Science and Technology of China, Radiation Oncology Key Laboratory of Sichuan Province, Chengdu 610041, China}

\author{Da Zhang}
\affiliation{Sichuan Cancer Hospital \& Institute, Sichuan Cancer Center, School of Medicine, University of Electronic Science and Technology of China, Radiation Oncology Key Laboratory of Sichuan Province, Chengdu 610041, China}

\author{Zhencen He}
\affiliation{Key Laboratory of Radiation Physics and Technology of Ministry of Education, Institute of Nuclear Science and Technology, Sichuan University, Chengdu 610064, China}
\affiliation{West China School of Basic Medical Sciences and Forensic Medicine, Sichuan University, Chengdu 610041, China}

\author{Zhimin Hu*}
\email{huzhimin@scu.edu.cn}
\thanks{The author is the corresponding author.}
\affiliation{Key Laboratory of Radiation Physics and Technology of Ministry of Education, Institute of Nuclear Science and Technology, Sichuan University, Chengdu 610064, China}

\date{\today}% It is always \today, today,

\begin{abstract}
Proton therapy has been rapidly advancing due to its excellent conformal index, but its relatively low relative biological effect (RBE) has somewhat limited its therapeutic efficacy for certain tumors. To address this, we previously proposed a nitrogen-targeting Proton-Carbon-Alpha-Therapy (Proton-CAT) enhancement method. In this letter, we present combined multi-scale DNA damage simulations and in vitro cell experiments, further investigating the mechanism of the Proton-CAT. It has been show that $^{15}$N enrichment significantly enhances complex DNA damage induced by high linear energy transfer(LET) particles within tumor regions. Under 30\% $^{15}$N conditions, $\alpha$ and $^{12}$C particle induced DSB++ increased by 175.19\% and 52.94\%, respectively. Furthermore, in vitro cell experiments using $^{15}$N-glutamine ($^{15}$N-Glu) as the $^{15}$N carrier indicated that high concentrations of $^{15}$N-Glu did not bring about significant cytotoxicity. Following 2 Gy irradiation, the cell viability in the 500 $\mu$g/mL $^{15}$N-Glu treated group exhibited a net reduction of about 15.41\% compared to the control group.This indicates that the enhanced effect of Proton-CAT primarily stems from increased complex DNA damage. This work provides a theoretical basis and multi-scale research framework for the development of the Proton-CAT.

\end{abstract}
\maketitle

Radiotherapy is an essential component of cancer treatment, with over half of cancer patients receiving radiotherapy\cite{delaney2005role}. Traditional photon radiotherapy typically results in dose deposition that decreases approximately exponentially along the direction of incidence, which tends to produce excessively high dose in the non‑target regions surrounding the tumor and may cause additional damage to normal tissues or organs \cite{prasanna2021normal,buzdar2009analysis}. In contrast, charged particle beams such as carbon ions or protons can achieve highly concentrated energy deposition at the end of their range due to the unique Bragg peak, thereby maximizing protection of critical organs\cite{ahmad2024particle, deng2026proton}. However, carbon ion therapy faces clinical limitations due to its high costs of construction and operation\cite{mohamad2017carbon,durante2017charged}. Conversely, proton therapy is widely adopted due to its mature technical infrastructure, yet its relatively low LET limits its effectiveness against certain types of tumors.\cite{paganetti2014relative}.
\begin{figure*}[t]
	\centering
    \hspace*{-10mm}
	  \includegraphics[width=0.86\textwidth]{Figures/Fig1.png}
	\caption{Conceptual diagram of the Proton-CAT model for MC simulation.
            (a) Schematic of the Proton-CAT nuclear reaction process under SOBP proton irradiation.
            (b) MC method for a macroscopic biological model composed of randomly rotating single-cells.
            (c) Single-cell MC model for calculating the DNA damage database.}
	\label{Fig1}
 \vspace*{-10pt}
\end{figure*}

% To overcome the biological limitations of proton therapy, several proton-enhancement strategies have been proposed in recent years. Current research has primarily focused on two directions. One approach has involved the use of high-atomic-number nanomaterials as radiosensitizers (such as AuNPs) to increase the generation of localized low-energy electrons and reactive oxygen species (ROS) through physical interactions\cite{mcmahon2011biological,kim2012enhanced}. The other approach has been based on nuclear reaction-driven proton capture therapy, such as proton boron capture therapy (PBCT), which has attracted significant attention because of its ability to generate high-LET $\alpha$ particles\cite{cirrone2018first,yoon2014application}. However, these existing approaches still face numerous challenges in clinical translation. For nanoparticles, complex cellular uptake mechanisms result in poor pharmacokinetics (PK) and low particle accumulation in tumor regions\cite{schuemann2020roadmap}; for PBCT, limitations in the reaction cross-section lead to poor energy matching with the Bragg peak of clinical protons, which restricts its therapeutic efficacy at specific depths\cite{tran2023current}.
Therefore, while preserving the advantages of proton therapy equipment, how to achieve high-LET therapeutic effects comparable to those of carbon‑ion therapy within the tumor target volume has become an urgent problem to be solved in the field of radiotherapy.\cite{mcintyre2023systematic,paganetti2025nrg}.

To address the limitations of existing proton therapy. we proposed a proton-enhanced irradiation method using $^{15}$N-labeled glutamine (C$_5$H$_1$$^{15}$N$_2$O$_3$, $^{15}$N-Glu) as a targeting carrier, termed nitrogen-targeting Proton-Carbon-Alpha Therapy (Proton-CAT) \cite{sun2025proton,wu2026spread,he2025effect}, as shown in Fig.~\ref{Fig1}(a). The Proton-CAT is based on the reaction
$p + ^{15}_7\mathrm{N} \rightarrow ^{16}_8\mathrm{O}^{*} \rightarrow ^{12}_6\mathrm{C} + \alpha + 5.0~\mathrm{MeV}$,
through which high-LET $\alpha$ and $^{12}$C secondary particles are generated in situ in the target region. Using this method, Proton-CAT allows protons to demonstrate high-LET therapeutic characteristics similar to those of heavy ions, thereby opening new avenues for integrating the physical advantages of proton therapy with the biological advantages of heavy‑ion therapy. In previous studies, we have demonstrated via Monte Carlo (MC) simulations that Proton-CAT effectively increases the yield of $\alpha$ and $^{12}$C secondary particles as well as the absorbed dose within the tumor region\cite{sun2025proton,wu2026spread}. However, how Proton-CAT induces complex DNA damage and enhances biological effects has not yet been systematically investigated. In this work, we further explore this issue through simulations and experiments.

Based on the research background described above, this work primarily used MC simulations to investigate the contribution of secondary particles generated by Proton-CAT to enhanced DNA damage. By establishing a single-cell DNA damage database and combining it with tissue-scale calculations, we quantified the contribution of these secondary particles to the induction of complex DNA damage in tumor regions. In addition, this work conducted in vitro cell irradiation experiments as a supplement to validate the simulation results, and to provide preliminary evidence for the biological effect and potential clinical feasibility of Proton-CAT.

At the microscopic level, a DNA damage database was established based on a single-cell model using the Density-Based Spatial Clustering of Applications with Noise (DBSCAN) algorithm. The G4EmDNAPhysics\_option2 and G4EmDNAChemistry\_option3 models were used to calculate DNA damage induced by proton, $\alpha$, and $^{12}$C ions within the cell nucleus over a range of incident energies \cite{incerti2018geant4,plante2017considerations,chatzipapas2023simulation}. At the macroscopic level, to simulate the extent of DNA damage in tumor tissue under different $^{15}$N enrichment conditions, three scenarios were established in the tumor region: a pure water model (100\% H$_2$O), 10\% $^{15}$N model (90\% H$_2$O + 10\% $^{15}$N), and 30\% $^{15}$N model (70\% H$_2$O + 30\% $^{15}$N). The relevant parameters are listed in Table.~\ref{Tab:Material}. To ensure computational accuracy and efficiency, the tumor region was modeled as a cube measuring $0.1 \times 0.1 \times 1.5~\mathrm{cm}^3$, and was flanked on both sides by normal tissue regions measuring $0.1 \times 0.1 \times 2~\mathrm{cm}^3$. As shown in Fig.~\ref{Fig1}(b), the cell units were arranged in a regular three-dimensional pattern within the tissue, and random three-dimensional rotations were applied to reduce the influence of geometric orientation on DNA damage statistics. A Spread-Out Bragg-Peak (SOBP) proton beam with an energy range of 47-65 MeV was used as the incident particle to ensure that the Bragg peak energy covered the entire tumor region. The calculations were performed using the G4HadronPhysics\_QGSP\_BIC\_AllHP model \cite{lu2021geant4,jarlskog2008physics}. The preconstructed single-cell DNA damage database was then combined with interpolation methods to estimate DNA damage in tumor tissue composed of cell units. Each simulation run employed 10$^8$ incident particles(relevant settings refer to \textcolor{blue}{Supplementary Material, Chapter I}).
\begin{table}[!htbp]
  \centering
   \vspace{-5mm}
  \caption{Atomic abundance and mass fraction of different materials in MC simulations}
  \begin{tabular*}{\columnwidth}{@{\extracolsep{\fill}}lcccccc@{}}
    \toprule
    \toprule
    \multirow{2}{*}{Material} & \multicolumn{3}{c}{Atomic abundance} & \multicolumn{3}{c}{Mass fraction} \\
    % \cmidrule(lr){2-4} \cmidrule(lr){5-7}
    & $^{1}_{1}\mathrm{H}$ & $^{16}_{8}\mathrm{O}$ & $^{15}_{7}\mathrm{N}$
    & $^{1}_{1}\mathrm{H}$ & $^{16}_{8}\mathrm{O}$ & $^{15}_{7}\mathrm{N}$ \\
    \midrule
    Pure water    & 66.7 & 33.3 & 0    & 11.2 & 88.8 & 0    \\
    10\% $^{15}$N & 60.0 & 30.0 & 10.0 & 8.8  & 69.5 & 21.7 \\
    30\% $^{15}$N & 46.7 & 23.3 & 30.0 & 5.4  & 42.9 & 51.7 \\
    \bottomrule
    \bottomrule
  \end{tabular*}
  \label{Tab:Material}
   	\vspace{-2mm}
\end{table}

The in vitro cell experiments were conducted using the human glioblastoma cell line U87 (U87-MG). U87-MG cells were seeded into a 96-well plates at a density of 6 $\times$ 10$^3$ cells per well. Following seeding, the cells were cultured for 24 h to ensure complete adherence. 24 h before irradiation, the cells were cultured in complete medium supplemented with different concentrations (0, 500, and 4000 $\mu$g/mL) of $^{15}$N-Glu. All experimental groups were performed in triplicate. The irradiation doses were 2, 4, and 6 Gy. Cell viability was determined using the CCK-8 assay(experimental procedures and cytotoxicity analysis refer to \textcolor{blue}{Supplementary Material, Chapter II}).
\begin{figure}[!htbp]
	\centering
     \hspace{-5mm}
	\includegraphics[width=0.45\textwidth]{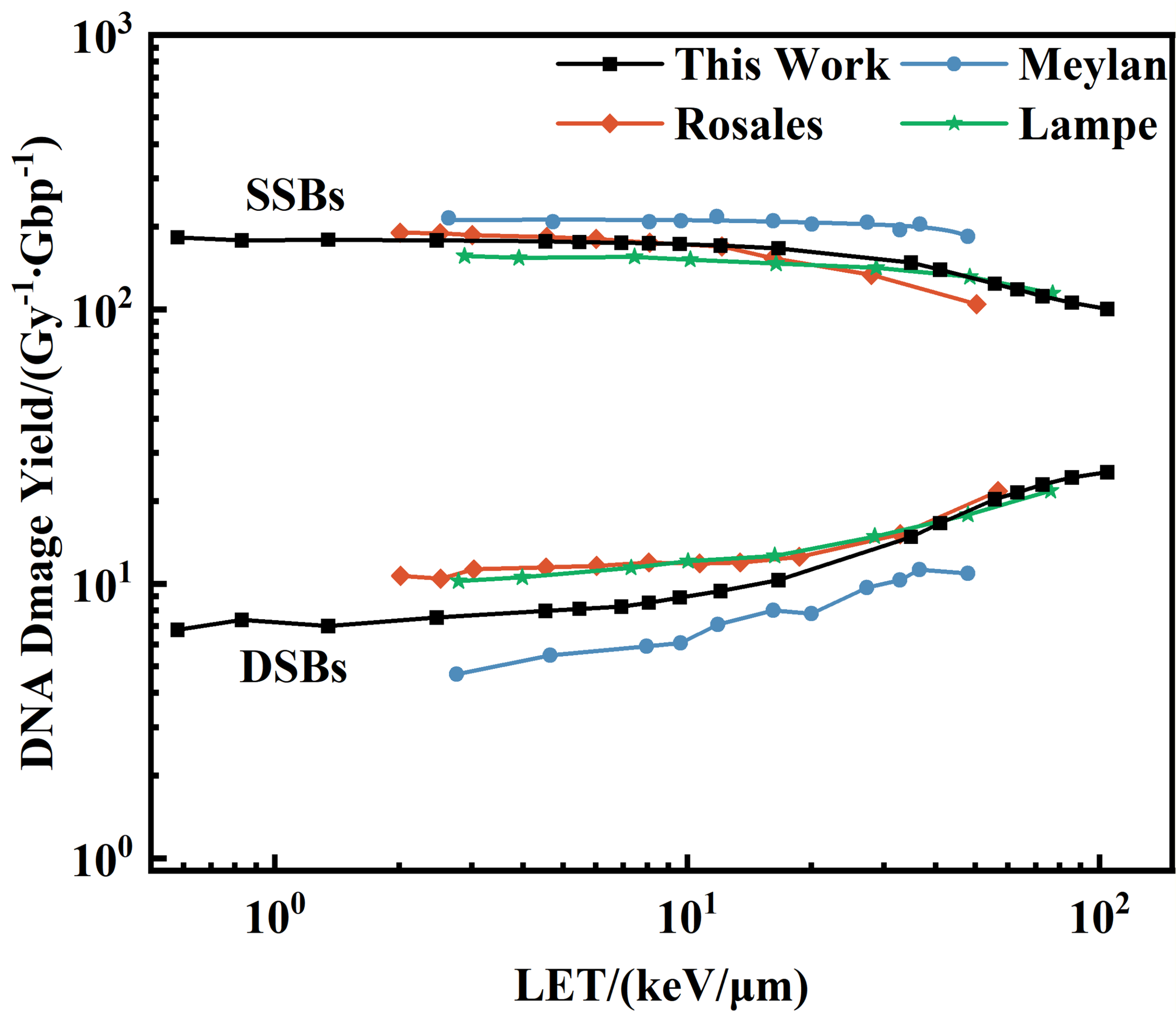}
	\caption{Comparison of calculation precision in DNA damage model.}
	\label{Fig2}
 \vspace*{-7pt}
\end{figure}

\begin{figure*}[!htbp]
	\centering
	\includegraphics[width=0.95\textwidth]{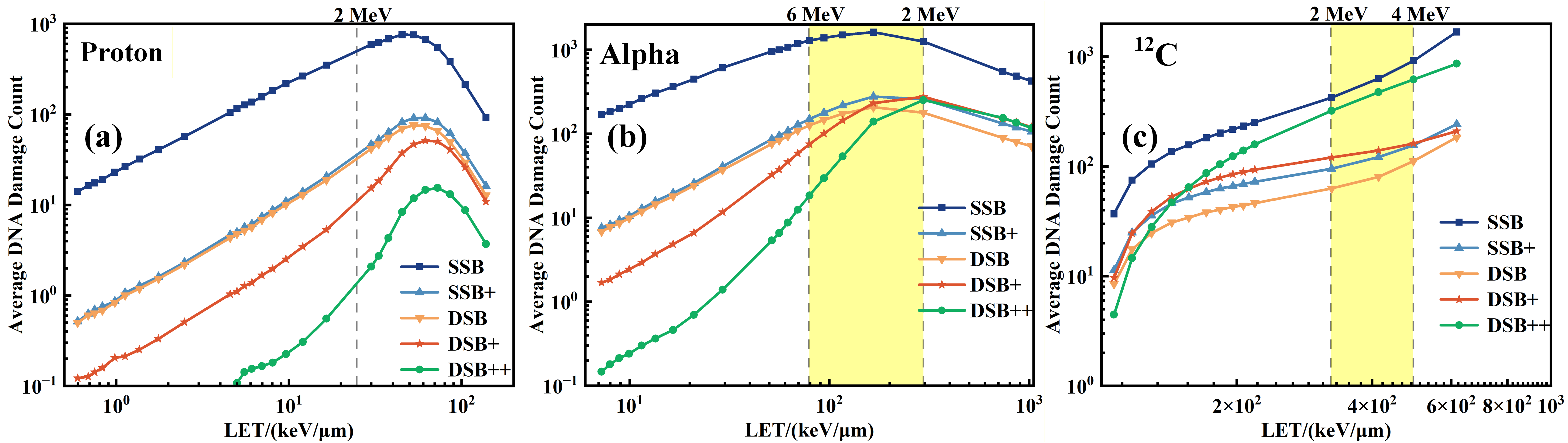}
	\caption{Calculation of the single-cell DNA damage database. (a) Proton. (b) $\alpha$ particle. (c) $^{12}$C particle. All DNA damage data has been normalized based on the total number of emitted particles in each case.}
	\label{Fig3}
 	\vspace{-5mm}
\end{figure*}

Fig.\ref{Fig2} shown a comparison of simulated results for single-strand breaks (SSB$_s$) and double-strand breaks (DSB$_s$) in fibroblast DNA under different proton irradiation energies with existing data\cite{meylan2017simulation, de2018accounting, lampe2018mechanistic}. Overall, they exhibited the same trend. The minor differences were primarily due to variations in parameter settings, such as DNA geometric models and DNA damage probabilities, rather than systematic errors introduced by the simulation method. Fig.\ref{Fig3} presents the yield statistics of various DNA damage types (SSB, SSB+, DSB, DSB+, DSB++) induced by proton, $\alpha$ particles, and $^{12}$C ions in glioma cells at different incident energies. The data indicated significant differences in DNA damage patterns induced by different particle types. Under identical incident energy conditions, low-LET protons primarily induce SSB$_s$, where as high-LET $\alpha$ particles and $^{12}$C ions are more likely to induce DSB$_s$. For instance, at 2 MeV, the numbers of DSBs induced by $\alpha$ particles and $^{12}$C ions were approximately 25.58 and 20.46 times, respectively, those induced by protons, and this disparity became greater with increasing particle energy.

Our previous studies have demonstrated that secondary particles produced in the $^{15}$N(p,$\alpha$)$^{12}$C nuclear reaction exhibit a broad energy distribution. About 45\% of the $\alpha$ particle energies fall within the 2-6 MeV range, while over 40\% of the $^{12}$C recoil nucleus energies are concentrated in the 2-4 MeV range\cite{sun2025proton}. This energy range corresponded to the interval in which elevated DSB$_s$ yields were observed in the present study. It indicates that even though the secondary particles are fewer in number than protons, their higher LET and energy distribution were more conducive to inducing complex DNA damage\cite{mladenova2022dna}. As a result, they still made a significant contribution to the overall biological effect in a mixed radiation field.

\begin{figure}
	\centering
      \hspace{-5mm}
	\includegraphics[width=0.5\textwidth]{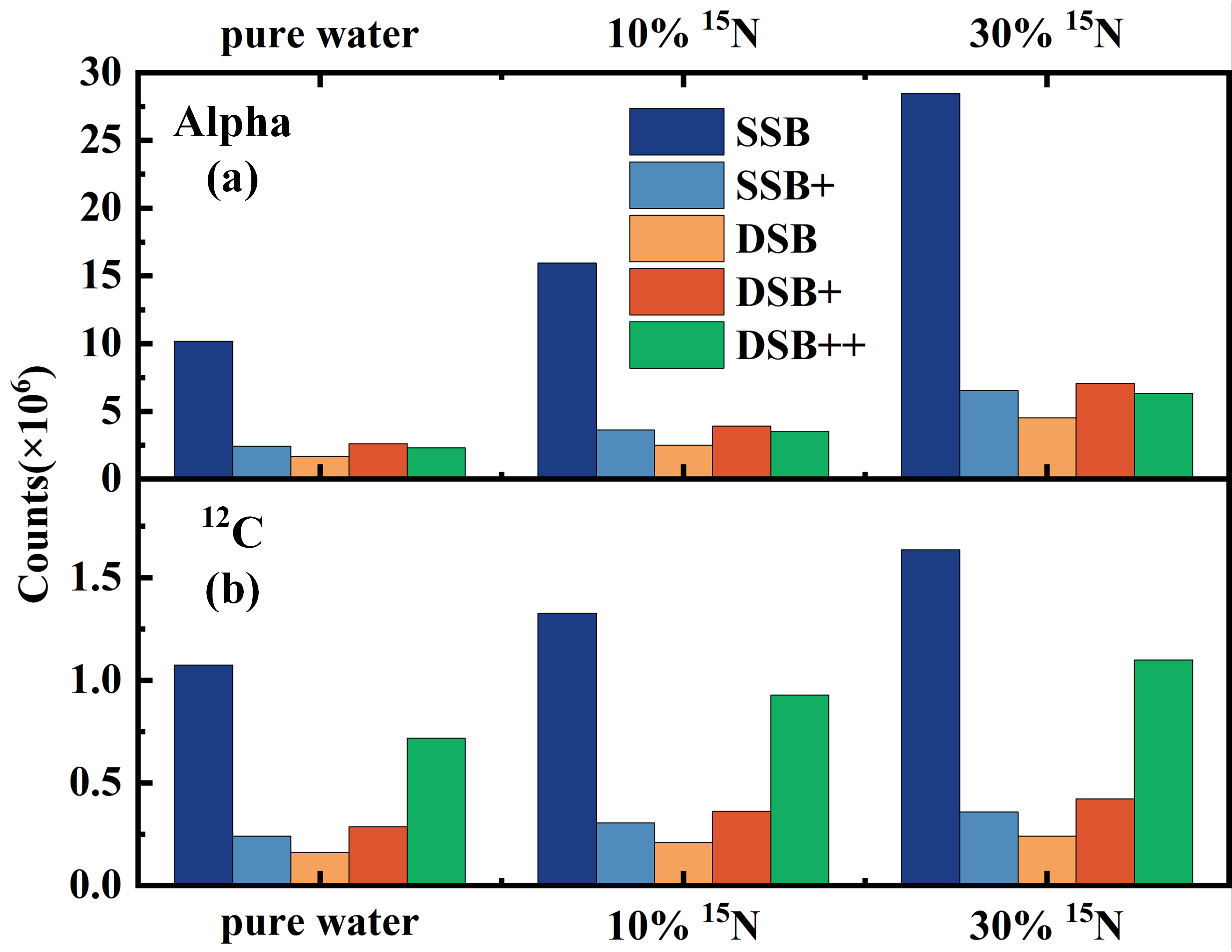}
	\caption{(a) Comparison of DNA damage types induced by $\alpha$ particle. (b) Comparison of DNA damage types induced by $^{12}$C particle.}
	\label{Fig4}
 	\vspace{-7mm}
\end{figure}

To evaluate the impact of the $^{15}$N(p,$\alpha$)$^{12}$C reaction in Proton-CAT on macroscopic tumor DNA damage, this work utilized a previously established DNA damage database and, in conjunction with the SOBP proton beam, performed simulation calculations of various types of DNA damage within the tumor region under different $^{15}$N enrichment conditions, as shown in Fig.~\ref{Fig4}. The results showed that the introduction of $^{15}$N significantly increased the incidence of various types of DNA damage induced by secondary particles within the tumor region. This enhancement became more pronounced as the $^{15}$N concentration increased. The contribution of secondary $\alpha$ particles is particularly significant.

\begin{figure*}[!htb]
  % \vspace{-\baselineskip}
	\centering
	  \includegraphics[width=\textwidth]{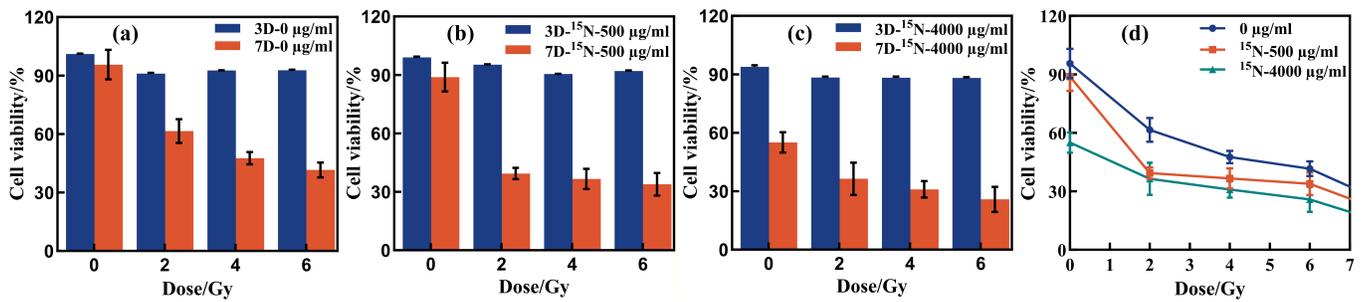}
	\caption{The time and dose dependent effects of $^{15}$N-Glu on the viability of proton irradiated U87-MG cells were evaluated using the CCK-8 assay.
            (a-c) The viability of U87-MG cells treated with different concentrations of $^{15}$N-Glu was measured at 3D and 7D after proton irradiation.
            (d) Dose-response curves of cell viability at 7D for different $^{15}$N-Glu concentrations.}
	\label{Fig5}
\end{figure*}

For SSB, compared to the pure water group, the increases under 10\% and 30\% $^{15}$N conditions were 56.93\% and 180.19\%, respectively, while the increase in DSB++ under 30\% $^{15}$N conditions reached 175.19\%. Although the number of DNA damage induced by $^{12}$C is relatively low, under 30\% $^{15}$N conditions, the induced SSB and DSB++ still increased by 52.29\% and 52.94\%, respectively. From a physical perspective, this enhancement stemmed from the high-LET secondary particles generated by the $^{15}$N(p,$\alpha$)$^{12}$C reaction. Compared with proton used in conventional proton therapy, $\alpha$ and $^{12}$C particles possess higher LET, capable of generating more intense primary ionization events and free radical production at the DNA level, leading to an increase in DNA damage yield. Notably, even in tissues lacking $^{15}$N, $\alpha$ and $^{12}$C particles can still be produced by nuclear reactions between protons and $^{16}$O, such as $^{16}\text{O}(p,\alpha)^{13}$N and $^{16}\text{O}(p,p'\alpha)^{12}\text{C}$, with threshold energies of approximately 5.66 MeV and 7.16 MeV\cite{chapman1967mechanism, chapman1967proton}, respectively. However, in tumor regions containing 10\% or 30\% $^{15}$N, the $^{15}$N(p,$\alpha$)$^{12}$C reaction became an additional source of secondary particles. This reaction significantly increased the production of secondary particles and led to higher yields of DNA damage induced by secondary particles than those in tumor regions lacking $^{15}$N.

Fig~\ref{Fig5} showed the changes in cell viability of U87-MG cells cultured at $^{15}$N-Glu concentrations of 500 $\mu$g/mL and 4000 $\mu$g/mL on day 3 (3D) and day 7 (7D) following proton beam irradiation. As shown in Fig.~\ref{Fig5}(a)-(c), on 3D after irradiation, the differences in cell viability among experimental groups were minor, with no significant changes observed regardless of $^{15}$N-Glu supplementation. However, by 7D, the disparities in cell viability between different treatment groups markedly amplified, exhibiting a pronounced time-dependent effect. Meanwhile, U87-MG cell viability decreased with increasing proton irradiation dose, showing a clear dose-dependent effect across the 2, 4, and 6 Gy groups.

Compared with the control group without $^{15}$N-Glu, cells pretreated with $^{15}$N-Glu exhibited significantly lower cell viability on 7D under the same dose. Under 2 Gy irradiation conditions, the net decrease in cell viability in the 500 $\mu$g/mL $^{15}$N-Glu-treated group reached approximately 15.41\% compared with the control group. As the dose or $^{15}$N-Glu concentration further increased, the apparent differences in cell viability among groups tended to level off. This may be because the CCK-8 assay primarily reflects the metabolic state of the cell population through mitochondrial dehydrogenase activity. It does not directly quantify cell proliferation or cell death\cite{khalef2024cell, adan2016cell}. Following proton irradiation, some damaged cells may remain in cell cycle arrest or undergo damage repair. At this stage, they may not yet undergoing lysis and death\cite{mirzayans2012new, fournier2004radiation}, and thus still capable of producing absorbance signals. This results in minimal changes in the measured viability rate. Despite this, the high-concentration $^{15}$N-Glu treated group maintained the lowest cell viability on 7D. This finding further supported the idea that the Proton-CAT sensitization mechanism did not induce immediate cell death, but instead impaired long-term proliferative capacity through cumulative DNA damage.

In summary, the physical basis of Proton-CAT lies in the nuclear reactions mediated by protons and $^{15}$N, which produce high-LET $\alpha$ and $^{12}$C particles. These secondary particles have shorter tracks and higher localized ionization densities, making them induce highly clustered ionization events and spatially correlated chemical damage at the DNA scale. Furthermore, in biological media, incident protons and their secondary particles generate a numerous low-energy secondary electrons along their trajectories. These secondary electrons can act directly on DNA\cite{ boudaïffa2000resonant}, while also forming reactive free radicals through the radiolysis of water molecules, which then migrate to key targets such as DNA and cause oxidative damage\cite{cadet1999hydroxyl}, thereby further amplifying local damage effects. Thus, the enhancement afforded by Proton-CAT does not merely equate to an increase in the macroscopic average dose. Instead, it leverages the intrinsic merits of proton therapy to synergistically boost the efficacy of biological damage through localized microscopic energy deposition and subsequent physicochemical processes.

In this work, we built on DNA damage simulation framework for the proposed Proton-CAT method by combining MC simulations with in vitro cell experiments. This approach was used to investigate its enhancement mechanism from both physical and biological perspectives. The results demonstrated that secondary particles generated by the $^{15}$N$(p,\alpha)^{12}$C reaction increased the complex DNA damage and enhanced the biological effects of proton therapy. This study revealed the physical basis of the Proton-CAT and clarified its multiscale relationship with complex DNA damage. These findings provided important support for the theoretical foundation and further development of Proton-CAT. However, the clinical translation of this method still requires further research. Future studies should focus on the delivery efficiency of $^{15}$N and related biological mechanisms to provide a scientific basis for the optimization and development of Proton-CAT.

\section*{Supplementary Material}
See the supplementary material for the MC simulation methods, simulation models and in vitro cell culture protocols in this work.

\begin{acknowledgments}
This work was supported by the Sichuan Science and Technology Program (Grant Nos. 2025NSFJQ0042 and 2025ZNSFSC0883), the Defense Industrial Technology Development Program (Grant No. JCKYS2023212808), the National Natural Science Foundation of China (Grant No. 12374234), Sichuan Province Special Fund for the Comprehensive Industrial Chain Development Project in Nuclear Medicine(Grant No. JG2025206).
\end{acknowledgments}

\section*{Data Availability Statement}
The data that support the findings of this work are available from the corresponding author upon reasonable request.

% \nocite{*}
\bibliography{aipsamp}% Produces the bibliography via BibTeX.
\end{document}